\documentclass[twocolumn,showpacs,preprintnumbers,amsmath,amssymb]{revtex4-1}
\usepackage{graphicx}
\usepackage{textcomp}

\begin{document}
\title{$\mathcal{PT}$ symmetry and anti-symmetry by anti-Hermitian wave coupling and nonlinear optical interactions}
  \normalsize
\author{Stefano Longhi}
\address{Dipartimento di Fisica, Politecnico di Milano and Istituto di Fotonica e Nanotecnologie del Consiglio Nazionale delle Ricerche, Piazza L. da Vinci
32, I-20133 Milano, Italy}

%
\bigskip
\begin{abstract}
\noindent
Light propagation in systems with anti-Hermitian coupling, described by a spinor-like wave equation, provides a general route for the observation of anti parity-time ($\mathcal{PT}$ ) symmetry in optics. Remarkably,  under a different definition of parity operator, a $\mathcal{PT}$ symmetry can be found as well in such systems. Such symmetries are ubiquitous in nonlinear optical interactions and are exemplified by considering modulation instability in optical fibers and optical parametric amplification. 
\end{abstract}

\maketitle


{\it Introduction.} Parity-time ($\mathcal{PT}$) symmetry is a very fruitful concept introduced in optics one decade ago, with a wealth of applications that are being explored in different areas of photonics and beyond \cite{r2,r3,r4,r5,r6,r7}. $\mathcal{PT}$ symmetry is typically found in optical media with a dielectric permittivity profile satisfying the symmetry condition $\epsilon(-\mathbf{r})=\epsilon^*(\mathbf{r})$, corresponding to spatial regions of balanced optical gain and loss. In such structures light propagation is described by a non-Hermitian Hamiltonian $\mathcal{H}$ which commutes with the parity-time operator $\mathcal{PT}$, i.e. $ [ \mathcal{H}, \mathcal{PT} ] =0$. A major property of $\mathcal{PT}$ systems is the existence of an abrupt phase transition, where the spectrum and structure of eigenvectors of $\mathcal{H}$ undergo a qualitative change and light behavior show intriguing properties \cite{r2,r3,r4}. Anti-$\mathcal{PT}$ symmetry, with the commutator replaced by the
anticommutator, or $ \{ \mathcal{PT} ,\mathcal{H} \} = 0$, represents a counterpart of $\mathcal{PT}$ symmetry. Anti-$\mathcal{PT}$ symmetry was suggested in optics in a few theoretical works \cite{r8,r9,r10,r11,r12} and demonstrated in a recent experiment based on flying atoms in a warm atomic-vapour cell \cite{r13}. As opposed to $\mathcal{PT}$ symmetry,  anti-$\mathcal{PT}$ symmetry is realized in optical media with a permittivity profile satisfying the antisymmetric relation $\epsilon^*(-\mathbf{r})=-\epsilon(\mathbf{r})$. Like for $\mathcal{PT}$-symmetric structures, anti-$\mathcal{PT}$-symmetric ones undergo a spontaneous phase transition of the eigenstates, which significantly changes light transport in the structure \cite{r11}.
    \begin{figure}[htb]
\centerline{\includegraphics[width=8.4cm]{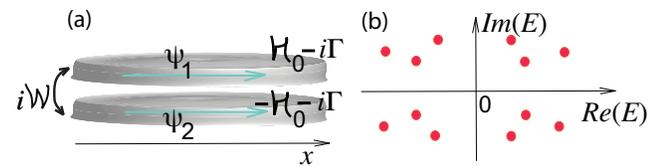}} \caption{ \small
(Color online) (a) Schematic of anti-$\mathcal{PT}$ symmetry based on anti-Hermitian coupling of two optical waves $\psi_1$ and $\psi_2$.  $\Gamma$ describes dissipation, $i \mathcal{W}$ anti-Hermitian coupling, and $\pm \mathcal{H}_0$  the free-evolution of the two waves in the absence of coupling and dissipation. (b) Symmetry properties of the energy spectrum of the Hamiltonian $\mathcal{H}$ for $\Gamma=0$. Anti-$\mathcal{PT}$ symmetry, $\mathcal{P^{\prime}T}$ symmetry and chiral $\mathcal{X}$ symmetry ensure that the spectrum is invariant under reflection with respect to both real and imaginary axes.}
\end{figure} 
 Optical systems with $\mathcal{PT}$ symmetry can show additional symmetries, such as chirality, charge conjugation or particle-hole symmetry \cite{r14,referee1,referee2,r15,r16}. $\mathcal{PT}$ and anti-$\mathcal{PT}$ symmetries are generally regarded as excluding one another. In fact, a $\mathcal{PT}$-symmetric system can be formally transformed into an  anti-$\mathcal{PT}$-symmetric one by Wick (imaginary time) rotation, i.e. by the transformation $ \mathcal{H} \rightarrow i \mathcal{H}$ \cite{r13,r18,r19}, which necessarily changes the physical structure of the medium. The definition of parity and time reversal operators, however, is not unique \cite{referee1,r20,r21}, so that is not physical forbidden for the same optical structure to introduce both kinds of symmetries. In this case, two distinct phase transitions can be observed, corresponding to an abrupt change of symmetries of system eigenvectors. So far, however, no examples of optical systems with $\mathcal{PT}$ symmetry and antisymmetry have been suggested. \\
 The aim of this Letter is to show that $\mathcal{PT}$ symmetry and antisymmetry are commonplace in spatially-extended spinor-like optical systems with anti-Hermitian wave coupling \cite{r11,r22,r23,r24}. Anti-Hermitian wave coupling is ubiquitous in nonlinear optical interactions, such as in optical parametric amplification, where symmetry breaking can be exploited for coherent pulse control. 
 
  {\it $\mathcal{PT}$-symmetry and anti-symmetry in anti-Hermitian coupled  waves.}  We consider two optical waves with spatial and/or temporal degrees of freedom that propagate in an optical structure and that are coupled via an anti-Hermitian interaction \cite{r11,r22,r23,r24}, as schematically shown in Fig.1(a). The temporal and/or spatial dynamics of the coupled waves is described by a spinor-like wave equation of the form
 \begin{equation}
 i \frac{\partial \psi}{\partial z}= \mathcal{H} \psi
 \end{equation} 
 where $\psi=(\psi_1,\psi_2)^T$, $\psi_1=\psi_1(x,z)$ and $\psi_2=\psi_2(x,z)$ are the wave amplitudes, $x$ is a continuous or discrete space/time variable that accounts for the degrees of freedom of the fields, and
 \begin{equation}
  \mathcal{H} \equiv \left( 
 \begin{array}{cc}
 \mathcal{H}_0-i {\Gamma} & i \mathcal{W} \\
 i \mathcal{W} & -\mathcal{H}_0-i {\Gamma}
 \end{array}
 \right)
 \end{equation}
 is the non-Hermitian Hamiltonian of the system. In Eq.(2),  $\mathcal{H}_0$, ${\Gamma}$ and $\mathcal{W}$ are self-adjoint operators that act on the variable $x$. Physically, the operators $\pm \mathcal{H}_0$ describe $^{\prime}$free-evolution$^{\prime}$ of each wave $\psi_1$ and $\psi_2$ in the absence of coupling and dissipation, $ i \mathcal{W}$ describes anti-Hermitian coupling of the two waves, whereas $\Gamma$ accounts for dissipation (including the one arising from non-Hermitian coupling \cite{r11,r22,r23}). The simplest case is obtained when  $\mathcal{H}_0$, ${\Gamma}$ and $\mathcal{W}$ are real numbers, i.e. $\mathcal{H}$ is a $2 \times 2$ matrix; in this limit the model (1) reduces to the dissipatively-coupled optical waveguide/resonator model introduced in Ref.\cite{r11}. 
In a two-component spinor-like wave equation, the definition of parity, time reversal and chiral (or charge conjugation) operators is not unique, and several possibilities have been discussed in \cite{referee1}; rather generally, such operators involve the combination of Pauli matrices $\sigma_k$ and the antiunitary complex conjugation operator $\mathcal{K}$. A first possibility is to introduce the parity $\mathcal{P}$ and time reversal $\mathcal{T}$ operators as follows
 \begin{equation}
 \mathcal{P}= \sigma_1 
  ,\; \; \mathcal{T}=\mathcal{I}\mathcal{K} 
 \end{equation}
 where $\mathcal{I}$ is the identity operator and $\mathcal{K}$ is the elementwise complex conjugation. It then readily follows that $\mathcal{H} \mathcal{PT}=-\mathcal{PT} \mathcal{H}$, i.e. $\{ \mathcal{H}, \mathcal{PT} \}=0$. Note that $\mathcal{T}$ is an antiunitary operator with $\mathcal{T}^2= \mathcal{I}$, $\mathcal{P}$ is a unitary linear operator with  $\mathcal{P}^2= \mathcal{I}$, and $[ \mathcal{T}, \mathcal{P}]=0$, i.e. $\mathcal{P}$ and $\mathcal{T}$ satisfy the general conditions of parity and time reversal operators \cite{r20,r25}. Also, since $\mathcal{PT}$ is antiunitary, $\mathcal{PT}$ antisymmetry can be viewed as particle-hole (or particle-antiparticle) symmetry for the non-Hermitian Hamiltonian $\mathcal{H}$ \cite{r24bis}, as earlier noted in \cite{referee1} in the $\mathcal{PT}$ symmetric context.
  Let us now assume that $\Gamma$ commutes with both $\mathcal{H}_0$ and $\mathcal{W}$, a condition which is generally met (for example whenever dissipation is uniform, i.e. independent of $x$). In this case, dissipation can be removed from the dynamics via the non-unitary transformation $\psi_{1,2}=\phi_{1,2} \exp(-\Gamma z)$, so that the spinor $(\phi_1, \phi_2)^T$ satisfies Eq.(1), where $\mathcal {H}$ is given by Eq.(2) with $\Gamma=0$. Therefore, in the following we can limit to consider the case $\Gamma=0$ in Eq.(2).  For $\Gamma=0$, besides anti-$\mathcal{PT}$ symmetry the coupled wave system exhibits two additional symmetries: chirality $\mathcal{X}$ (also referred to as charge symmetry in \cite{referee1}), $\mathcal{H} \mathcal{X}=-\mathcal{X} \mathcal{H}$, and $\mathcal{P}^{\prime} \mathcal{T}$ symmetry, $\mathcal{P}^{\prime} \mathcal{T} \mathcal{H}=\mathcal{H} \mathcal{P}^{\prime} \mathcal{T}$,  under the different definition of parity operator $\mathcal{P}^{\prime}$. Operators $\mathcal{X}$ and $\mathcal{P}^{\prime}$ are defined by 
\begin{equation}
\mathcal{X}=\sigma_2
  ,\; \; 
\mathcal{P}^{\prime}=\sigma_3 
\end{equation}  
  with $ \mathcal{P}^{\prime \; 2}=\mathcal{I}$ and $[ \mathcal{P}^{\prime}, \mathcal{T}]=0$. Note that the definition of parity operators $\mathcal{P}$, $\mathcal{P}^{\prime}$ and chiral (charge) symmetry $\mathcal{X}$ introduced in this work are different than those used in \cite{referee1}. Note also that, since $\mathcal{P}^{\prime} \mathcal{T}$ is antiunitary, $\mathcal{P}^{\prime} \mathcal{T}$ can be also viewed as a time reversal symmetry for $\mathcal{H}$ \cite{r24bis}. Ant-$\mathcal{PT}$, $\mathcal{P}^{\prime} \mathcal{T}$ and $\mathcal{X} $ symmetries of the wave equation implies some restrictions on the energy spectrum and symmetries of eigenfunctions of the Hamiltonian. Specifically, if $\psi$ is an eigenvector of $\mathcal{H}$ with eigenvalue $E$, then (i) $\mathcal{P} \mathcal{T} \psi$ is an eigenvector of $\mathcal{H}$ with eigenvalue $-E^*$; (ii) $\mathcal{P}^{\prime} \mathcal{T} \psi$ is an eigenvector of $\mathcal{H}$ with eigenvalue $E^*$; (iii) $\mathcal{X} \psi$ is an eigenvector of $\mathcal{H}$ with eigenvalue $-E$. The energy spectrum $E$ of $\mathcal{H}$ is thus invariant under specular reflections with respect to both real and imaginary axis, as schematically shown in Fig.1(b). While the chiral symmetry is always in the unbroken phase, the $\mathcal{PT}$ antisymmetry is unbroken if $E=-E^*$, i.e. when the energy spectrum of $\mathcal{H}$ is entirely imaginary, whereas the $\mathcal{P^{\prime}T}$ symmetry is unbroken if $E^*=E$, i.e. if the energy spectrum of $\mathcal{H}$ is entirely real.  
 To investigate the phase transitions arising from symmetry breaking, let us assume that $\mathcal{W}$ commutes with $\mathcal{H}_0$, and let us indicate by $\varphi_0$ an eigenvector of $\mathcal{H}_0$ with eigenvalue $E_0$, i.e. $\mathcal{H}_0 \varphi_0= E_0 \varphi_0$ and $\langle \varphi_0 | \varphi_0 \rangle=1$. It then readily follows that the spinors
  \begin{equation}
  \psi_{\pm}= \left( 
  \begin{array}{c}
  i \langle \varphi_0 | \mathcal{W} \varphi_0 \rangle \\
  E_{\pm}-E_0
  \end{array}
  \right) \varphi_0
  \end{equation}
  are eigenvectors of $\mathcal{H}$ with eigenvalues
  \begin{equation}
  E_{\pm}= \pm \sqrt{E_0^2 -\langle \varphi_0 | \mathcal{W} \varphi_0 \rangle^2}.
  \end{equation}
  In the most common cases of anti-Hermitian coupling of spatially or temporally extended optical waves, such as in the case of nonlinear optical interactions discussed below, the energy spectrum of $\mathcal{H}_0$ is unbounded either from above or below, i.e. $|E_0|$ can take arbitrarily large values, whereas the coupling strength $\langle \varphi_0 | \mathcal{W} \varphi_0 \rangle$ remains bounded and vanishes uniformly as the anti-Hermitian coupling $\mathcal{W}$ vanishes. Therefore, from Eq.(6) it follows that the energy spectrum $E$ always contains real eigenvalues, i.e. anti-$\mathcal{PT}$ symmetry breaking is {\it thresholdless}. On the other hand, if $E_0=0$ is not an eigenvalue of $\mathcal{H}_0$, i.e. if the uncoupled waves do not admit zero-energy modes, for a sufficiently small value of coupling the system is in the unbroken $\mathcal{P}^{\prime} \mathcal{T}$ phase. 
  
  {\it  Anti-$\mathcal{PT}$ and $\mathcal{PT}$ symmetries in nonlinear optical interactions.} Nonlinear optical interactions provide an important and experimentally accessible route to realize anti-Hermitian wave coupling. In a nonlinear medium, two optical waves with different color or polarization state  are anti-Hermitian coupled when mixed with a third (usually strong) wave. Earlier works highlighted that either anti-$\mathcal{PT}$ \cite{r10} or $\mathcal{PT}$ \cite{r26,r27,r28} symmetries can be introduced in nonlinear optical interactions, however such previous works focused to some special optical settings and did not recognize the simultaneous presence of both symmetries. 
 Here we show that $\mathcal{PT}$ and anti-$\mathcal{PT}$ symmetries are ubiquitous in nonlinear optical interactions.\\
 As a first example, let us consider the phenomenon of modulational instability (MI) in a nonlinear optical fiber \cite{r29,r30}. MI can be explained as a thresholdless symmetry breaking and the existence of a zero-energy mode. Wave propagation in a lossless optical fiber is described by the nonlinear
Schr\"odinger equation \cite{r29,r30}
 \begin{equation}
 i \frac{\partial \mathcal{E}}{\partial z}=-\frac{k^{\prime\prime}}{2} \frac{\partial^2 \mathcal{E}}{\partial x^2}+\frac{\omega n_2}{2c} | \mathcal{E}|^2 \mathcal{E}
 \end{equation}
  where  $\mathcal{E}=\mathcal{E}(x,z)$ is the complex electric field envelope, $\omega$ is the angular frequency, $c$ is the speed of light,  $z$ is the propagation distance along the fiber axis, $x=t-z/v_g$ is the retarded time, $v_g=(\partial k / \partial \omega)^{-1}$ is the group velocity, $n_2$ is the Kerr coefficient ($n_2>0$ for a fiber) and $k^{\prime\prime}= (\partial^2 k / \partial \omega^2)$ is the group velocity dispersion.  As is well known, if a continuous-wave field with amplitude $\mathcal{E}_0$ is injected into the fiber, a temporal MI can arise from noise in the negative dispersion regime  $k^{\prime\prime}<0$. Interestingly, MI can be viewed as a symmetry breaking process. In fact, let us look for a solution to Eq.(7) in the form $\mathcal{E}(x,z)= \mathcal{E}_0 (1+ \delta \mathcal{E}) \exp(-i \sigma z)$, where $\sigma \equiv \omega n_2 |\mathcal{E}_0|^2 /(2 c)$ and $\delta\mathcal{E}=\delta\mathcal{E}(x,z)$ is a small perturbation. After setting  $\delta\mathcal{E}=\psi_1(x) \exp(-i Ez)-i \psi_2^*(x) \exp(iEz)$, it readily follows that $(\psi_1, \psi_2)^T$ and $E$ are the eigenvectors and corresponding eigenvalues of the Hamiltonian $\mathcal{H}$, defined by Eq.(2), with $\mathcal{W}= \sigma$, $\Gamma=0$ and $\mathcal{H}_0=-(k^{\prime\prime} /2) (d^2/dx^2)+\sigma$. The eigenfunctions and corresponding eigenvalues of $\mathcal{H}_0$ are given by $\varphi_0(x)=\exp(i \Omega x)$ and $E_0=\sigma+k^{\prime\prime} \Omega^2/2$, where $\Omega$ is the temporal frequency of the perturbation. Since the spectrum of $\mathcal{H}_0$ is unbounded, the anti-$\mathcal{PT}$ symmetry is always in the broken phase. This means that, for sufficiently large frequency detuning $\Omega$, the eigenvalues $E_{\pm}$, given by Eq.(6), are real and the corresponding eigenvectors $(\psi_1,\psi_2)^T$ are not eigenvectors of the $\mathcal{PT}$ operator.  On the other hand, the $\mathcal{P^{\prime}T}$-symmetry can be in the unbroken phase (for $k^{\prime\prime}>0$, i.e. in the positive dispersion regime) or in the broken phase (for $k^{\prime\prime}<0$, i.e. in the negative dispersion regime). In the former case the eigenvalues $E_{\pm}$ are real at any frequency $\Omega$ and MI is prevented: this is because  the Hamiltonian $\mathcal{H}_0$ does not have a zero-energy mode. On the other hand, in the latter case $\mathcal{H}_0$ shows a zero-energy mode at frequency $\Omega_m=\sqrt{-2 \sigma / k^{\prime\prime}}$, and the $\mathcal{P^{\prime}T}$-symmetry breaking is thresholdless. The frequency $\Omega_m$ of the zero-energy mode corresponds to the frequency sideband of perturbations with the highest MI gain.\par
  As a second example, let us consider parametric amplification of short optical pulses at carrier frequencies $\omega_1$ (signal wave) and $\omega_2$ (idler wave) in a lossless $\chi^{(2)}$ nonlinear medium pumped by a strong and nearly continuous-wave  pump at frequency $\omega_3=\omega_1+\omega_2$; Fig.2(a). Indicating by $k=k(\omega)= \omega n(\omega) /c$ the linear dispersion relation of the dielectric medium and $x=t- k^{\prime}_3 z$ the retarded time, where $1/k^{\prime}_3=(\partial k / \partial \omega)_{\omega_3}^{-1}$ is the group velocity of the pump wave, the coupled equations that describe parametric amplification of the signal and idler fields in undepleted pump regime read \cite{r31,r32}
  \begin{eqnarray}
  i \frac{\partial A_1}{\partial z}  =  \left[ k_1-k(\omega_1+i \partial_x)+ik^{\prime}_3 \partial_x \right] A_1-\Delta_1 A_2^* \exp(i \Delta k z) \\
  i \frac{\partial A_2^*}{\partial z}  =  -\left[ k_2-k(\omega_2+i \partial_x)-ik^{\prime}_3 \partial_x \right] A_2^*+\Delta_2^*  A_1 \exp(-i \Delta k z) 
  \end{eqnarray}
  where $A_1=A_1(x,z)$ and $A_2=A_2(x,z)$ are the amplitudes of signal and idler waves, respectively, $k_l=k(\omega_l)$ and $n_l=n(\omega_l)$ are the wave number and refractive index at frequency $\omega_l$ $(l=1,2,3)$, $A_3=A_3(x)$ is the amplitude of the incident strong pump wave, $\Delta k= k_3-k_2-k_1$ is the wave vector mismatch, $\Delta_{1,2}= k_{1,2}\chi^{(2)}_{eff}A_3/(2n_{1,2}^2)$,
  and $\chi^{(2)}_{eff}$ is the effective nonlinear dielectric susceptibility. Note that in writing Eqs.(8) and (9) all orders of material dispersion are considered. The most usual form of Eqs.(8) and (9) is obtained by expanding the dispersion relation $k(\omega_l+i \partial_x)$ up to second order, i.e. by assuming $k(\omega_l+i \partial_x)=k_l+i k_l^{\prime} \partial_x-(1/2) k^{\prime \prime}_l \partial^2_x$,  so that pulse walk off and group velocity dispersion are considered in first-order approximation \cite{r30}. Coupled equations (8) and (9) can be viewed as a wave equation for the two-component spinor variable $\psi=(A_1, A_2^*)^T$, whose Hamiltonian $\mathcal{H}$ however differs from the canonical form (2). In this case $\mathcal{P^{\prime}T}$ and anti-$\mathcal{PT}$ symmetries are hidden and can be revealed after a non-local transformation of wave variables. In fact, let us assume that the pump wave is nearly monochromatic and not chirped, so that without loss of generality one can set $A_3=i \bar{A}_3$ with $\bar{A}_3$ real and slowly-varying function of $x$. Let us then introduce the following integral transformation for signal and idler wave amplitudes
  \begin{eqnarray}
  A_1(x,z)  =  -\exp \left( i \frac{\Delta k z}{2} \right)   \int d \xi \mathcal{G}(x- \xi,z) \psi_1(\xi,z)  \\
  A_2(x,z)  =   \sqrt{\frac{k_2}{k_1}} \frac{n_1}{n_2} \exp \left( i \frac{\Delta k z}{2} \right) \int d \xi \mathcal{G}^*(x- \xi,z) \psi_2^*(\xi,z) 
  \end{eqnarray}
  where $\mathcal{G}(\tau,z)$ is given by 
  \begin{equation}
  \mathcal{G}(\tau,z)  = \frac{1}{2 \pi} \int d \Omega \exp \left[ -i \Omega \tau+i \beta(\Omega) z \right]  
  \end{equation}
  and where we have set
  \begin{equation}
 2  \beta(\Omega)=k_2-k_1+k(\omega_1+\Omega)-k(\omega_2-\Omega)-2 k^{\prime}_3 \Omega.
  \end{equation}
and $\mathcal{G}(\tau,0)=\delta(\tau)$.The explicit form of the kernel $\mathcal{G}$ depends on the dispersion curve of the optical medium. For example, if $k(\omega_1 +\Omega)$ and $k(\omega_2 -\Omega)$ entering in Eq.(13) are expanded up to second order in $\Omega$, which is a reasonable approximation for not too short optical pulses, one readily obtains
\begin{equation}
\mathcal{G}(\tau,z)=\sqrt{\frac{i}{\pi z(k_1^{\prime \prime}-k_2^{\prime \prime}) } } \exp \left\{ - i\frac{\left[ z \frac{k_1^{\prime}+k_2^{\prime}-2k_3^{\prime}}{2}-\tau \right]^2}{z (k_1^{\prime \prime} -k_2^{\prime \prime})} \right\}
\end{equation}
In terms of the transformed amplitudes $\psi_1$ and $\psi_2$, it can be shown that the evolution equation of the spinor $( \psi_1, \psi_2)^T$  is given by Eq.(1), where the Hamiltonian $\mathcal{H}$ is given by Eq.(2) with $\Gamma=0$ and with 
\begin{equation}
\mathcal{H}_0  = \mathcal{H}_0(i \partial_x) \; , \;\; 
\mathcal{W} = \sqrt{k_1k_2}  \chi^{(2)} \bar{A}_3/(2 n_1 n_2)
\end{equation}
where we have set
\begin{equation}
 \mathcal{H}_0( \Omega) = (1/2) \left[ k_3-k( \omega_1+\Omega) -k(\omega_2-\Omega) \right].
\end{equation}
Interestingly, the Hamiltonian $\mathcal{H}_0$ corresponds to the phase mismatch curve, so that a zero-energy mode $\mathcal{H}_0(\Omega)=0$ indicates that perfect phase matching can be realized at some spectral frequency $\Omega$. Note also that, within the quadratic approximation of the dispersion relation, one has
\begin{equation}
\mathcal{H}_0  \simeq  (1/2) \left[  \Delta k-i (k_1^{\prime}-k_2^{\prime}) \partial_x+ (1/2)(k_1^{\prime \prime} +k_2^{\prime \prime}) \partial^2_x\right].
\end{equation}
For a continuous-wave pump, the eigenvalues and corresponding eigenvectors of $\mathcal{H}_0$ are given by $E_0=H_0(\Omega)$ and $\varphi_0= \exp(-i \Omega x)$. The energy spectrum $E_0$ is generally unbounded, so that anti-$\mathcal{PT}$ symmetry is likely to be always in the broken phase. However, if phase matching is broadband and the spectrum of incident signal/idler pulses is sufficiently narrow, anti-$\mathcal{PT}$ symmetry is unbroken [Fig.2(b)]. Physically, this means that all spectral components of the signal and idler waves are amplified as they propagate along the nonlinear medium, which shows an effective $^{\prime}$constant refraction$^{\prime}$ \cite{r11,r13}. An interesting case is obtained when the phase matching curve is spectrally flat; in the parabolic approximation a flat dispersion curve is achieved when the signal and idler waves have the same group velocity $k_1^{\prime}=k_2^{\prime}$ but opposite group velocity dispersion $k_1^{\prime \prime}=-k_2^{\prime \prime}$ \cite{r33}. From Eq.(17) it then follows that $\mathcal{H}_0 \simeq \Delta k/2$ is flat. This means that $\mathcal{PT}$ antisymmetry and  $\mathcal{P^{\prime}T}$ symmetry are exchanged  at $\mathcal{W}=|\Delta k |/2$. At such a pump level symmetry breaking corresponds to a second-order exceptional point with global non-Hermitian eigenvector collapse, i.e. at any frequency component $\Omega$ of generated signal and idler photons the parametric amplification process arises from the coalescence of pairs of eigenvalues and corresponding eigenvectors of the Hamiltonian. Such a global collapse of eigenvectors at the exceptional point realizes a $^{\prime}$photonic catastrophe$^{\prime}$ for optical pulses, similar to the one recently studied in Ref.\cite{r34}:  near the symmetry breaking transition parametric amplification is strongly sensitive to changes of initial pulse excitation condition, which could be exploited for coherent pulse tailoring. In fact, let us assume $k^{\prime}_2=k^{\prime}_1$, $k^{\prime \prime}_2=-k^{\prime \prime}_1$, a pump level  $\mathcal{W}= \Delta k /2$ (symmetry breaking point), and let us indicate by $f(x)$ and $g(x)$ the signal and idler pulses that excite the crystal at $z=0$, i.e. 
$A_1(x,0)=f(x)$ and $A_2(x,0)=g(x)$. After setting 
\begin{equation}
\theta(x) \equiv f(x)-i(n_2/n_1) \sqrt{k_1/k_2}\; g^*(x)
\end{equation}
 and indicating by $\theta(x,z)$ the solution to the  dispersive wave equation $i \partial_z \theta=(k_1^{\prime \prime}/2) \partial_x^2 \theta$ with the initial condition $\theta(x,0)=\theta(x)$, it can be readily shown that  
the signal and idler waves  asymptotically behave as $A_1(x,z) \sim -i \mathcal{W}z \exp(i \Delta k z/2)  \theta(x-\Delta k^{\prime}z,z)$ and 
$A_2(x,z) \simeq - \mathcal{W} z  \sqrt{k_2/k_1}(n_1/n_2) \exp(i \Delta k z/2) \theta^*(x-\Delta k^{\prime}z,z)$, where $\Delta k^{\prime}=k^{\prime}_1-k^{\prime}_3$. Physically, this means that the spreading and pulse waveform of the signal and idler waves are asymptotically determined by the interference $\theta(x)$ of initial pulse shapes $f(x)$ and $g(x)$. The coherent superposition propagates as a dispersive wave with a group velocity dispersion $k^{\prime \prime}_1$. In particular, if the injected pulses are tailored such that their interference $\theta(x)$ is much shorter than $f(x)$ and $g(x)$, the temporal analogue of $^{\prime}$superdiffraction$^{\prime}$ \cite{r34}, i.e. $^{\prime}$superdispersion$^{\prime}$,  could be observed: the signal/idler optical pulses spread in time at a rate much higher than the one observed in the absence of wave coupling.\\
The opposite regime is realized when phase matching  is largely mismatched over a broad frequency spectrum [Fig.2(c)]. In this case $\mathcal{PT}$ antisymmetry is broken while $\mathcal{P^{\prime}T}$ is unbroken. In the parabolic approximation, this condition is met, for example, when signal and idler pulses travel at the same group velocity ($k_1^{\prime}=k_2^{\prime}$) and the signs of mean group velocity dispersion $(k_1^{\prime \prime}+k_1^{\prime \prime})/2$ and of phase mismatch $\Delta k$ are opposite.\par

 \begin{figure}[htb]
\centerline{\includegraphics[width=8.4cm]{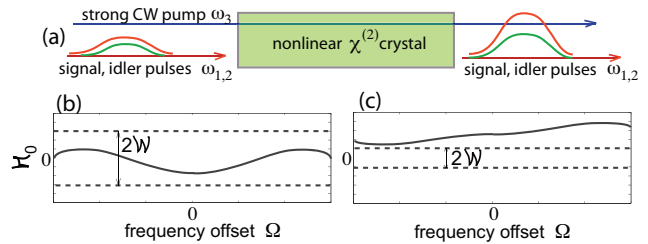}} \caption{ \small
(Color online) (a) Schematic of parametric amplification of weak signal/idler optical pulses by a strong CW pump. (b,c) Phase matching curve, showing the behavior of $\mathcal{H}_0(\Omega)$. In (b)  anti-$\mathcal{PT}$ symmetry is in the unbroken phase (broadband phase matching), whereas in (c)  the $\mathcal{P^{\prime}T}$ symmetry is in the unbroken phase (large broadband phase mismatch).}
\end{figure} 

 {\it Conclusions.} Optical systems possessing $\mathcal{PT}$ or anti-$\mathcal{PT}$ symmetries have attracted a great interest in recent years. While $\mathcal{PT}$  symmetry has been investigated and experimentally observed in different optical setting, anti-$\mathcal{PT}$ symmetry has been so far observed only using atomic vapor systems \cite{r13}. A general belief is that such symmetries exclude one another, since they require different symmetries for the dielectric permittivity. Here we have shown that, under different yet consistent definitions of parity operators, $\mathcal{PT}$ symmetry and antisymmetry can be simultaneously introduced for anti-Hermitian coupled optical waves, such as in nonlinear optical wave mixing and optical parametric amplification. Our results indicate that nonlinear optical interactions could provide a useful laboratory tool for the experimental realization of $\mathcal{PT}$ symmetry and antisymmetry in optics with potential applications to coherent pulse control.

\end{document}